\input{template.tex}
\begin{document}
\begin{titlepage}
\vspace{-.2in}
\begin{flushright}
UICHEP-TH/93-9\\
UdeM-LPN-TH-93-172
\end{flushright}
\begin{center}
{\large {\bf Inter-Relations of Solvable Potentials} }\\

\vspace{1 in}

Asim Gangopadhyaya$^{(a)}$,
Prasanta K. Panigrahi$^{(b)}$\footnote{Address after Oct. 1,
Department of Physics, Univ. of Hyderabad, Hyderabad, A.P., India, 500134.}
and Uday P. Sukhatme$^{(c)}$
\end{center}
\begin{tabular}{l l}
a) &Department of Physics, Loyola University Chicago,
Chicago, IL 60626;\\
b) &Laboratoire de Physique Nucl\'eaire, Universit\'e de
Montr\'eal, Montr\'eal, \\ &Qu\'ebec, Canada H3C3J7;\\
c) &Department of Physics, University of Illinois at
Chicago,\\& 845 W. Taylor Street, Chicago, IL 60607-7059.\\
\end{tabular}
\vspace{1.0in}
\begin{abstract}
Solvable Natanzon potentials in
nonrelativistic quantum mechanics are known to group into two disjoint classes
depending on whether the Schr\"odinger equation can be reduced to a
hypergeometric or a confluent hypergeometric equation. All the
potentials within each class
are connected via point canonical transformations. We establish a
connection between the two classes with appropriate  limiting procedures
and redefinition of parameters, thereby inter-relating all known solvable
potentials.
\end{abstract}

\end{titlepage}
\newpage
It is well known that the Natanzon potentials\cite{natanzon} are exactly
solvable
in nonrelativistic quantum mechanics.
These potentials are of two types corresponding to
whether the Schr\"{o}dinger equation can be reduced to either a hypergeometric
or a confluent hypergeometric equation.
Those that lead to a
hypergeometric equation (confluent hypergeometric equation) will be called
type-I (type-II) potentials. It has been shown\cite{cooper,de,levai}
that the members within each class can be mapped
into each other by point canonical transformations (PCT);
however, members of these two different classes cannot be connected by a PCT.
Since a hypergeometric differential equation
reduces to a confluent hypergeometric one under appropriate limits,
it is reasonable to expect that the potentials of the above mentioned two
classes can also be connected by a similar procedure. The purpose of
this note is to establish a connection between specifically
chosen potentials in each class. A convenient choice is
the so called shape invariant potentials\cite{gendenshtein} which
form a distinguished class in
the sense that their spectra can be determined entirely
by an algebraic procedure,
akin to that of the harmonic oscillator, without ever referring
to the underlying differential equations.
We provide a list of mappings that connect shape invariant
type-I potentials to type-II potentials.

Before proceeding further, it is worth reviewing point
canonical transformations in nonrelativistic quantum mechanics.
We consider a time-independent Schr\"{o}dinger equation
with a potential function  $V(\alpha_i; x)$ that depends
upon several parameters $\alpha_i$ (we will use $\hbar =
2m = 1 $):
\begin{equation}
\left[ -{d^2 \over {dx^2}} + V(\alpha_i; x) -
E(\alpha_i) \right] \psi(\alpha_i; x) = 0.
\label{eq.1}
\end{equation}
Under a point canonical transformation which  replaces the independent
variable $x$ by $z$  $(x=f(z))$ and transforms the wave function
$\left[ \psi(\alpha_i; x)
=v\left( z\right) \tilde{\psi}  \left(
\alpha_i; z \right) \right]$,
the Schr\"{o}dinger equation transforms into:
\begin{equation}
-{d^2 \tilde{\psi} \over {dz^2}}
-\left\{ {2v' \over v} - {f'' \over f'} \right\}
{d \tilde{\psi} \over {dz}}
+ \left\{
 f'^2 \left[ V(\alpha_i; f(z)) - E(\alpha_i) \right] +
\left( {f''v' \over f'v} - {v' \over v}  \right)
\right\} \tilde{\psi}
= 0.
\label{eq.2}
\end{equation}
Requiring the first derivative term to be absent
gives $v(z)=C\sqrt{f'(z)}$.
This then leads to another Schr\"{o}dinger equation with a new potential.\\
\begin{equation}
\left[ -{d^2 \over {dz^2}}
+ \left\{
 f'^2 \left[ V(\alpha_i; f(z)) - E(\alpha_i) \right] +
{1 \over 2}
\left( {3f''^2 \over 2f'^2} - {f''' \over f'}  \right)
\right\}
\right] \tilde{\psi} \left( \alpha_i; z \right) = 0.
\label{eq.3}
\end{equation}\\
In general, this is not an eigenvalue equation, unless
$\left\{ f'^2 \left( V(\alpha_i; f(z)) - E(\alpha_i)\right) \right\}$
has a term independent of
$z$, which will act like the energy term for the new
Hamiltonian.   This condition constrains allowable
choices for the function $f(z)$.  For a general potential
$V(\alpha_i; f(z))$, many choices for $f(z)$ are
still possible that would give rise to
Schr\"odinger type eigenvalue equations, and thus, if we
have one solvable model, we can generate many others from
it.

Ref.\cite{de} contains a list of functions $f(z)$
that relate all shape invariant potentials of type-I (type-II) to the Scarf
(harmonic oscillator) potential. In the following, we will
present two examples where suitable limits
take one beyond class barriers, and connect type-I
potentials to those of type-II. In particular, we shall exhibit the
limiting procedures that convert (a) the Scarf potential into
the harmonic oscillator potential; and (b) the generalized P\"oschl-Teller
into either the Morse or the harmonic oscillator potentials.
In Table I,
we provide additional examples of limiting procedures and redefinition of
parameters.

\vspace{.25in}

\noindent {\bf Scarf potential to harmonic oscillator:}\\
The Scarf potential, given by
$$V_{Scarf}(x)=-A^2 + (A^2+B^2-A\alpha){\rm sec}^2(\alpha x)-
B(2A-\alpha){\rm tan}(\alpha x){\rm sec}(\alpha x)$$
goes into the three-dimensional harmonic oscillator potential (HO)
$$V_{HO} = {1 \over 4} \omega^2 r^2 +
{l(l+1) \over r^2} - \left(l+{3 \over 2}\right)\omega$$
after a shift of origin $x\rightarrow \left(r-{\pi \over 2\alpha}\right)$,
a redefinition of parameters
${A}\rightarrow \left( {\omega \over \alpha} + \alpha {(l+1) \over 2}
\right),\;\;
{B}\rightarrow \left( {\omega \over \alpha} - \alpha {(l+1) \over 2}
\right)$, and then taking the limit $\alpha\rightarrow 0.$

\vspace{.25in}

\noindent {\bf Generalized P\"oschl-Teller potential to Morse:}\\
The generalized P\"oschl-Teller potential (GPT)
$$V_{GPT}(r) = A^2 + (A^2+B^2+A\alpha) \,
{\rm cosech}^2 \, (\alpha r+\beta) -
B(2A+\alpha) {\rm cosech} \, (\alpha r+\beta)
{\rm coth} \, (\alpha r+\beta)$$
can be converted into two shape invariant potentials of type-II by taking
appropriate limits. One obtains the Morse potential when
$B\rightarrow {1 \over 2}
B e^\beta, \;{\rm and \;one \;takes \;the \;limit} \;
\beta\rightarrow\infty. $
Alternatively, one gets the three dimensional harmonic oscillator potential
when\cite{cheung}
$$A\rightarrow
\left({\omega \over \alpha}-\alpha {(l+1)\over 2}\right),
\; B\rightarrow \left({\omega \over \alpha}+
\alpha {(l+1)\over 2}\right),\;\alpha\rightarrow 0, \; \beta\rightarrow 0.$$

Here, it is worth noting that we have given straightforward
routes for going from type-I to type-II potentials.
Type-I potentials give rise to hypergeometric differential
equation which has three regular singular points.  Two of
them merge in the limiting procedures stated above,
and as expected one gets a confluent hypergeometric equation.
The reverse procedure of going from type-II to type-I is not
well defined.
We also provide a figure with information on different
limiting procedures and point canonical transformations that
take type-I potentials among each other or reduce them to
type-II potentials.

One of us (PKP) would like to thank the Physics Department of the University
of Illinois for warm hospitality, where this work began.
This work was supported in part by the U.S. Department of
Energy under grant number DE-FG02-84ER40173 and by the
National Science and Engineering Research Council
(NSERC) of Canada.


\newpage
\noindent
{\bf TABLE CAPTIONS:}\hspace {.5in} Table I \\

\noindent
Limiting procedures
and redefinition of parameters that relate
type-I to type-II potentials.
\newpage

\noindent
{\bf FIGURE CAPTIONS:}\hspace {.5in} Figure I \\

\noindent
Limiting procedures and
point canonical transformations that take type-I
potentials among each other or reduce them to
type-II potentials.
Potentials are same as those in Ref.\cite{de}.
\textheight 10.0in
\newpage
\thispagestyle{empty}
\begin{tabular}{|l|l|l|}  \hline
{\bf Type-I Potential} &  {\bf Type-II Potential } &
{\bf Limits }\& {\bf Redef.}\\
& &      {\bf of Parameters}					\\
\hline
{\bf Generalized P\"oschl-Teller }  &  {\bf Harmonic Oscillator }	&
$A \rightarrow \left[
{\omega \over \alpha} - \alpha \left({l+1 \over 2}\right)\right]$	\\
$V(r)=A^2+{(A^2+B^2+A\alpha)\over {\rm sinh^2}(\alpha r+\beta)}$  	&
$V(r)={1\over 4} \omega^2r^2 + { l(l+1) \over r^2}-(l+{3\over 2})\omega $&
$B \rightarrow \left[
{\omega \over \alpha} +  \alpha \left({l+1 \over 2}\right)\right]$
\\
$\;\;-{B(2A+\alpha){\rm coth}(\alpha r+\beta) \over
{ {\rm sinh}(\alpha r+\beta)} }$ &
$0<r<\infty,\;\;\;\;E_n=2n\omega $  					&
$\alpha \rightarrow 0,\; \beta\rightarrow 0$\\
\cline{2-3}
& {\bf Morse Potential } &
$A \rightarrow A $							\\
$-\beta <\alpha r<\infty$						&
$V(x)=A^2 + B^2e^{-2\alpha\,x}$						&
$B \rightarrow {Be^\beta \over 2}$	\\
$E_n=A^2-\left(A-n\alpha\right)^2$					&
$- 2B \left(A+{\alpha \over 2}\right) e^{-\alpha\,x}$ &
$r \rightarrow x$\\
$A<B$ & $-\infty<x<\infty $ &
$\beta\rightarrow \infty$			\\
&$E_n=A^2-\left(A-n\alpha\right)^2$ &\\
\hline
{\bf Scarf } &  {\bf Harmonic Oscillator }&     			\\
$V(x)= -A^2+ {(A^2+B^2-A\alpha ) \over
{{\rm cos^2}(\alpha x)} } $  		&
$V(r)={1\over 4} \omega^2r^2 + { l(l+1) \over r^2}-(l+{3\over 2})\omega$      &
$A \rightarrow \left[
{\omega \over \alpha} + \alpha \left({l+1 \over 2}\right)\right]$	\\
$\;\;-{B(2A-\alpha){\rm tan}(\alpha x) \over
{{\rm cos}(\alpha x)}  }$	&
$0<r<\infty$								&
$B \rightarrow \left[
{\omega \over \alpha} -  \alpha \left({l+1 \over 2}\right)\right]$	\\
$-{\pi \over 2\alpha}<x<{\pi \over 2\alpha},\; A>B,$			&
$E_n=2n\omega $ 							&
$x\rightarrow r+{\pi \over 2\alpha}$\\
$E_n=\left(A+n\alpha\right)^2-A^2$ & &
$\alpha \rightarrow 0\; $						\\
\hline
{\bf Scarf (Hyperbolic)}  &  {\bf Morse Potential } 		&	\\
$V(x)=A^2+{(-A^2+B^2-A\alpha)\over {\rm cosh^2}(\alpha x+\beta)}$  	&
$V(x)=A^2 + B^2e^{-2\alpha\,x}$						&
$A\rightarrow A$\\
$\;\;+{B(2A+\alpha){\rm tanh}(\alpha x+\beta) \over
{ {\rm cosh}(\alpha x+\beta)} }$ 					&
$- 2B \left(A+{\alpha \over 2}\right) e^{-\alpha\,x}$ 			&
$B\rightarrow - {Be^\beta \over 2}$					\\
$-\infty<x<\infty, \;A>0 $ 	& $-\infty<x<\infty $ &
$\beta\rightarrow \infty$			\\
$E_n=A^2-\left(A-n\alpha\right)^2$ & $E_n=A^2-\left(A-n\alpha\right)^2$	& \\
\hline
{\bf Eckart}& {\bf Coulomb} & 						\\
$V(r)=A^2+{B^2 \over A^2} - 2B {\rm coth}\; \alpha r$			&
$V(r)=- {e^2 \over r} + { l(l+1) \over r^2} $	&
$A\rightarrow \alpha (l+1)$						\\
$+A(A-\alpha){\rm cosech}^2\alpha r $&
$\;\;\;\;\;\;\;\;\;+{e^4 \over 4(l+1)^2}$				&
$ B \rightarrow {\alpha \over 2} e^2$\\
$0<r<\infty,\; B>A^2,\; A>0$ 						&
$0<r<\infty $ 	&	$\alpha\rightarrow 0$				\\
$E_n= A^2-\left(A+n\alpha\right)^2$&&\\
$+{B^2 \over A^2} -{B^2 \over {(A+n\alpha)}^2}$ &
$E_n={e^4 \over 4}\left( {1\over (l+1)^2}-{1\over (n+l+1)^2}\right)$ &\\
\hline
{\bf Rosen-Morse I}& {\bf Coulomb} & 					\\
$V(x)=-A^2+{B^2 \over A^2} + 2B {\rm tan} \;\alpha x$			&
$V(r)=- {e^2 \over r} + { l(l+1) \over r^2} $	&
$A\rightarrow \alpha (l+1)$						\\
$+A(A-\alpha){\rm sec}^2\alpha x $					&
$\;\;\;\;\;\;\;\;\;+{e^4 \over 4(l+1)^2}$ 				&
$ B \rightarrow {\alpha \over 2} e^2$					\\
$-{\pi \over 2\alpha} < x <{\pi \over 2\alpha} $ 			&
$0<r<\infty $ 	&
$x\rightarrow r-{\pi \over 2\alpha}$ 					\\
$E_n=- A^2+\left(A+n\alpha\right)^2$&&$\alpha\rightarrow 0$		\\
$+{B^2 \over A^2} -{B^2 \over {(A+n\alpha)}^2}$ &
$E_n={e^4 \over 4}\left( {1\over (l+1)^2}-{1\over (n+l+1)^2}\right)$ &\\
\hline
\end{tabular}\\

\begin{center}
{\bf Table I}
\end{center}
\end{document}